\documentclass[epj]{svjour}

\usepackage{graphicx}

\renewcommand{\l}{\left}						
\renewcommand{\r}{\right}						
\newcommand{\degr}{^\circ}						

\begin{document}

\title{Emergence of semi-localized Anderson modes in a disordered photonic crystal as a result of overlap probability}
\author{A. R. Hashemi \and M. Hosseini-Farzad \and Afshin Montakhab\mail{montakhab@shirazu.ac.ir}
}                     
%
%
\institute{Department of Physics, College of Sciences, Shiraz University, Shiraz 71454, Iran}
\date{\today}
%
\abstract{
In this paper we study the effect of positional randomness on transmissional properties of a two dimensional photonic crystal as a function of a randomness parameter $\alpha$ ($\alpha=0$ completely ordered, $\alpha=1$ completely disordered). We use finite-difference time-domain~(FDTD) method to solve the Maxwell's equations in such a medium numerically. We consider two situations: first a $90\degr$ bent photonic crystal wave-guide and second a centrally pulsed photonic crystal micro-cavity. We plot various figures for each case which characterize the effect of randomness quantitatively. More specifically, in the wave-guide situation, we show that the general shape of the normalized total output energy is a Gaussian function of randomness with wavelength-dependent width. For centrally pulsed PC, the output energy curves display extremum behavior both as a function of time as well as randomness. We explain these effects in terms of two distinct but simultaneous effects which emerge with increasing randomness, namely the creation of semi-localized modes and the shrinking (and eventual destruction) of the photonic band-gaps. Semi-localized (i.e. Anderson localized) modes are seen to arise as a synchronization of internal modes within a cluster of randomly positioned dielectric nano-particles. The general trend we observe shows a sharp change of behavior in the intermediate randomness regime (i.e. $\alpha \approx 0.5$) which we attribute to a similar behavior in the underlying overlap probability of nano-particles.
} 
%
\titlerunning{Emergence of semi-localized modes in a disordered photonic crystal ...}
\maketitle

\section{Introduction} \label{sec1}
Photonic crystals~(PC), structures with a periodic refractive index distribution, are a subject of intense worldwide research~\cite{r01,r02,r03,Singh20085083}. Among various PC structures, two dimensional~(2D) PCs based on a finite cluster of cylinders of infinite length have attracted most attention~\cite{r01,r02}, since these structures are relatively easy to fabricate or simulate. PCs have potential for many technological applications. For example, in contrast with the traditional wave-guides in which guiding principle is based on total internal reflection, PC wave-guides have a fundamentally different mechanism~\cite{r04}. These wave-guides can propagate the light in a path on the scale of wavelength without diffraction losses even at corners, because light leakage is forbidden for frequencies in the photonic band-gap~\cite{r05}.

The effect of disorder in 2D PCs with $90\degr$ bent wave-guide was investigated by Langtry et al.~\cite{r06} as well as Kwan et al.~\cite{r07}. In the former, the sensitivity of wave-guiding to different degree of disorder, mostly in radius of cylinders, refractive index, and filling factor have been studied by highly accurate multipole method. In the latter, by using multiple-scattering method, it has been shown that the transmission spectrum of the wave-guide is more sensitive to the uniformity in position of first layer surrounding the wave-guide.

In the first part of this article, we study, using finite-difference time-domain (FDTD) method~\cite{TafloveFDTD}, normalized output energy of a $90\degr$ bent wave-guide in 2D PC consisting of a square array of long dielectric cylinders in the air background. By introducing a wide range of randomness in the position of all cylinders, we show that the output energy normalized to the ordered case have a Gaussian shape, when plotted with respect to randomness factor. This behavior is shown for any wavelength in the original PC band-gap.

In the second part of this article, we study the effect of positional randomness on the light energy output from a PC micro-cavity, where four cylinders are removed from the center of square array and a short Ricker wavelet pulse~\cite{Schneider98} is initiated in the center of the micro-cavity. We use standard FDTD method to study the time evolution of output energy for TM electromagnetic field in the above system. The normalized output energy curve as a function of time for different randomness factors starts out with a sharp minimum followed by a broad maximum. Previous investigations are mainly concerned with cavity mode frequencies and quality factor versus strength of randomness~\cite{r09} as well as the variation of density-of-states with respect to reduced frequency~\cite{r10}. Additionally, some researchers have undertaken theoretical studies of positional disorder effect on resonant mode distributions and resonance frequencies in 2D photonic crystal micro-cavity~\cite{r11,r12}, and experimental activity accompanied by numerical simulation of spatial intensity distribution of localized modes in 2D open microwave cavity~\cite{r13}. In the latter, for detection and simulation of localized modes, the random medium is excited by more than one thin antennas around the center of randomized PC. However, in the present work, we excite the medium by only one central Ricker wavelet pulse and then the system is left to its own in order to study how the energy of the pulse exit from boundaries of PC when it is subjected to the different strengths of randomness.

Using transfer matrix method, the first study of the effect of induced randomness in radius or refractive index of the cylinders have shown the appearance of propagation states in the original band-gap~\cite{r14}. Band-gaps, although with some narrowing, were found to persist even for large amount of disorder. This phenomenon has been observed both theoretically~\cite{r15,r16,r17,r18} as well as experimentally~\cite{r19,r20}, especially in the case of positional disorder in cylinders~\cite{r15,r21}. These investigations mainly concerned analysis of transmission through PC slabs. However, in the present work the source is placed in the center of our system and band-gap resistance to disorder is observed in the transmission spectrum of the output light. We show that the formation of non-central semi-localized modes, besides the light localization in the central micro-cavity, can explain dynamical behavior of normalized output energy for various randomness factors.

Here, the emergence of semi-localized (or Anderson localized) modes \cite{and.loc} in our randomized PC is due to synchronization of internal reflection modes in nano-particles which are situated near each other in a suitable yet random manner. Previously, similar ideas have been proposed in the context of coupled particle cavities~\cite{Lichmanov98,Briskina02} in order to interpret experimental results in powder random lasers~\cite{Markushev90,Briskina96}. However, the medium there is active and the particle sizes (4-20 $\mu$m) are much larger than emission wavelength. Here, on the other hand, we see such effects arises naturally as a result of randomness in a passive medium within clusters of nano-particles ($\simeq$ 100 nm). More recently, Anderson localization has become a controversial topic in disordered PC's \cite{and.loc,Nature200962,PhysRevLett.99.253901}. Here, we provide further evidence for existence of such modes in 2D PC's. Similar localization phenomena has also been seen in \cite{PhysRevE.71.026612} where randomness in size of cylinders has been considered.

This paper is arranged as follows: in Section~\ref{sec2} we briefly explain the set-up for our numerical simulations. Section~\ref{sec3} entails the main results of our study which includes numerical results as well as discussion of the relevant physical effects. We summarize our main results in Section~\ref{sec4}.

\section{Numerical Simulation}\label{sec2}
\begin{figure}[htbp!]
    \includegraphics[width=\columnwidth]{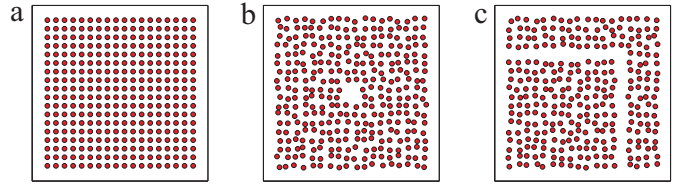}
    \caption{(a) Perfect photonic crystal, (b) disordered PC ($\alpha = 0.5$) with cavity, (c) disordered PC ($\alpha = 0.5$) with wave-guide.}
    \label{fig:fig1}
\end{figure}

In this paper we concentrate on a PC model which consist of a square lattice of cylinders with refractive index 3 with radius $0.3d$, where $d$ is the lattice constant of the crystal [Fig.~\ref{fig:fig1}(a)]. The surrounding of cylinders is air with refractive index 1. This kind of crystal has a band-gap in wavelength region $ 3.0d < \lambda < 3.7d $~\cite{Lang2004}. In order to introduce disorder in the position of cylinders, we consider displacement $\delta d$ for each cylinder, which is a random number between zero and a percentage of lattice constant $d$. We call this percentage \emph{randomness factor} $\alpha$. We consider the cylinders axes aligned in $z$ direction, so in the disordered case, like Fig.~\ref{fig:fig1}(b), each cylinder displaces from its ordered position, as $\delta d_x = u_x \times \alpha \times d$ in $x$ direction and as $\delta d_y = u_y \times \alpha \times d$ in $y$ direction; where $u_x$ and $u_y$ are random numbers with flat distribution between -1 and +1. We perform our simulation with $d = 180$ nm and so, we choose visible light, $\lambda \sim 550$ nm, in the edge of band-gap, to study the effect of randomness better.

To simulate propagation of light, we use FDTD method, and for truncating computational space, we use CPML absorbing boundary condition~\cite{CPML2000}.

Our model is symmetric along $z$ axis, so choosing TMz mode, we carry out the simulation in two dimensions. For each randomness factor $\alpha$, we repeat simulation with several different random arrangements (ensemble) of cylinders, and average over the various realizations of randomness. For first scenario, i.e. propagation of light by a wave-guide in PC [Fig.~\ref{fig:fig1}(c)], we enter a plane-wave to the wave-guide and calculate the output energy at the other end; while for the second scenario, i.e. propagation of a pulse from the center of PC, we use a Ricker wavelet pulse with central wavelength, $\lambda = 550$ nm, as initial pulse and calculate total output energy on all sides of crystal. In the second scenario, we remove four central cylinders to prevent the source from overlapping with randomly arranged neighboring cylinders and we therefore have a cavity in the center of the PC [Fig.~\ref{fig:fig1}(b)].

\section{Numerical Results and Discussions}\label{sec3}
\subsection{wave-guide PC}
\begin{figure}[htbp!]%
	\includegraphics[width=\columnwidth]{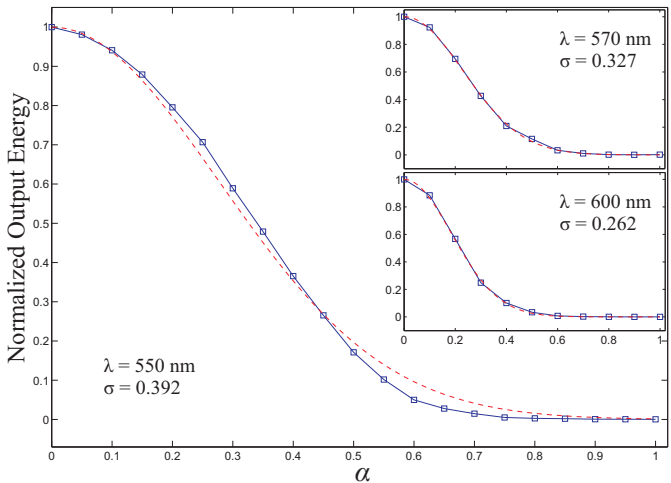}
	\caption{Output energy of guided light by a $90\degr$ bent wave-guide in the photonic crystal (solid line) compared with a Gaussian function with $\sigma = 0.392$ (dashed line). Upper inset shows the output for $\lambda = 570$ nm compared with a Gaussian with $\sigma = 0.327$, and lower inset shows the output for $\lambda = 600$ nm compared with a Gaussian with $\sigma = 0.262$. The axis on the inset are the same as the main figure.}
    \label{fig:fig2}
\end{figure}

Figure~\ref{fig:fig2} shows normalized output of guided energy in a $90\degr$ bent wave-guide PC. For each randomness factor, the output energy is averaged over 200 different random samples and normalized to the output energy of the ordered case ($\alpha=0$). As we expect, output energy decreases with increasing disorder, however, we note the following observations:
\begin{list}{}{}
    \item[\emph{i})] The general shape of the output energy curve seems to fit well with a Gaussian function, $E_{out}=E_{order} e^{-\alpha^2/\sigma^2}$. Increasing number of samples used in averaging makes an increasingly better fit with such function.
    \item[\emph{ii})] We observe that the general Gaussian shape is independent of the wavelength of light we used [Fig.~\ref{fig:fig2} insets]. However, the width of the Gaussian ($\sigma$) decreases with increasing $\lambda$, whether $\lambda$ lies in the edge of band-gap or not.
    \item[\emph{iii})] The relatively large value of $\sigma$ indicates certain robustness to disorder in this wave-guiding effect. For example, in Fig.~\ref{fig:fig2}, at 15\% randomness, output energy is still about 90\% of the ordered amount.
\end{list}
It is obvious that the above observations is related to the relative robustness of the photonic band-gap structure for small and intermediate values of $\alpha$ and its eventual destruction for large $\alpha$ [see Fig.~\ref{fig:fig7}]. However, the Gaussian shape of the figure is a curious effect, a point which we will come to later on this paper.

\subsection{centrally pulsed PC}
\begin{figure}[htbp!]
    \includegraphics[width=\columnwidth]{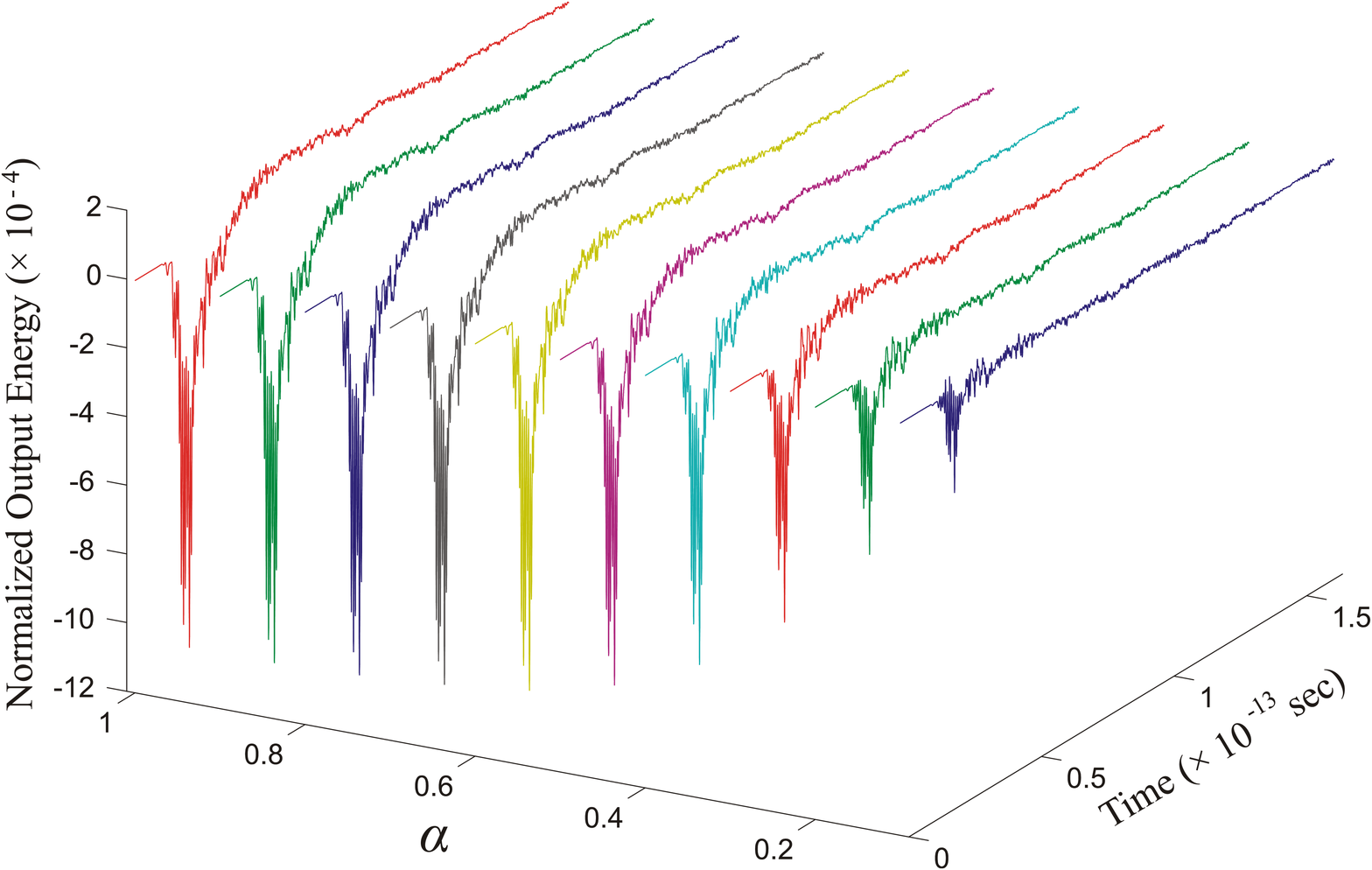}
    \caption{Normalized output energy of crystal as a function of time for different randomness factors. The output energy of the ordered crystal has been subtracted.}
    \label{fig:fig3}
\end{figure}

We now turn our attention to the second scenario where a Ricker wavelet source is placed in the middle of the PC and the total output is monitored along all edges. To study the changes in localized modes in the crystal, we plot output energy of pulse from the crystal as a function of time for different randomness factors in Fig.~\ref{fig:fig3}. For each $\alpha$, we average output energy over 300 different random samples and subtract output curve of ordered crystal ($\alpha = 0$) from it, then normalize the result to the total input energy. Therefore, in Fig.~\ref{fig:fig3}, each curve represents the (average) effect of a given randomness ($\alpha$) on the relative output energy as a function of time. An important characteristic stands out: each curve starts out with a sharp minimum followed by a broad maximum before setting into a constant zero value.

\begin{figure}[htbp!]
    \includegraphics[width=\columnwidth]{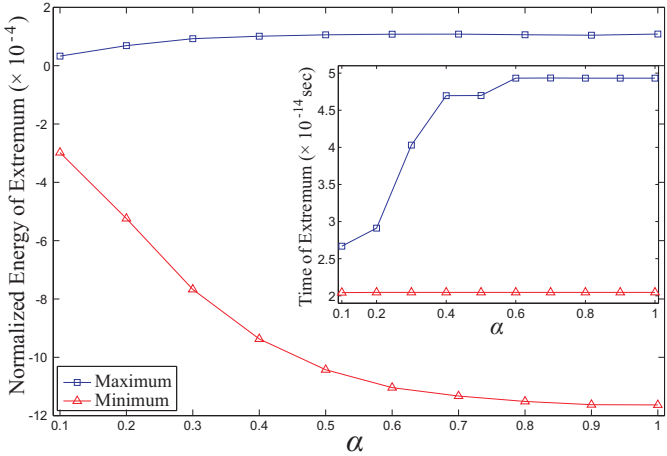}
    \caption{Energy of maximum and minimum of Fig.~\ref{fig:fig3} versus randomness factor. Inset shows time of occurrence of the maximum and minimum. Note that the maximum is insensitive to $\alpha$ while the minimum decreases continuously with $\alpha$. On the other hand, the time of occurrence of the minimum is insensitive to $\alpha$ while the peak of the output shifts with increasing disorder.}
    \label{fig:fig4}
\end{figure}

As is shown in Fig.~\ref{fig:fig4}, the (absolute) value of the minimum increases with increasing $\alpha$, while the peak of the broad maximum is largely unaffected by the randomness. On the other hand, the minimum occurs at a fixed time regardless of $\alpha$, while the maximum occurs at longer times with increasing $\alpha$ (inset Fig.\ref{fig:fig4}).

It is easy to understand the behavior in Fig.~\ref{fig:fig4} by considering the effect of randomness on the PC. The minimum corresponds to the leaving of the unaffected modes from the crystal, i.e. the modes outside the band-gap thus leaving the crystal straight away without impedance. Therefore the constant time of occurrence for this minimum can easily be understood.

As randomness increases the band-gap shrinks, so we expect that a larger portion of the modes leave the crystal without being affected; but we see that the minimum is lowered by randomness, which means, the unaffected modes decreases by increasing randomness. The root of this apparent discrepancy is in the creation of the so-called semi-localized modes.

\begin{figure}[htbp!]
    \includegraphics[width=\columnwidth]{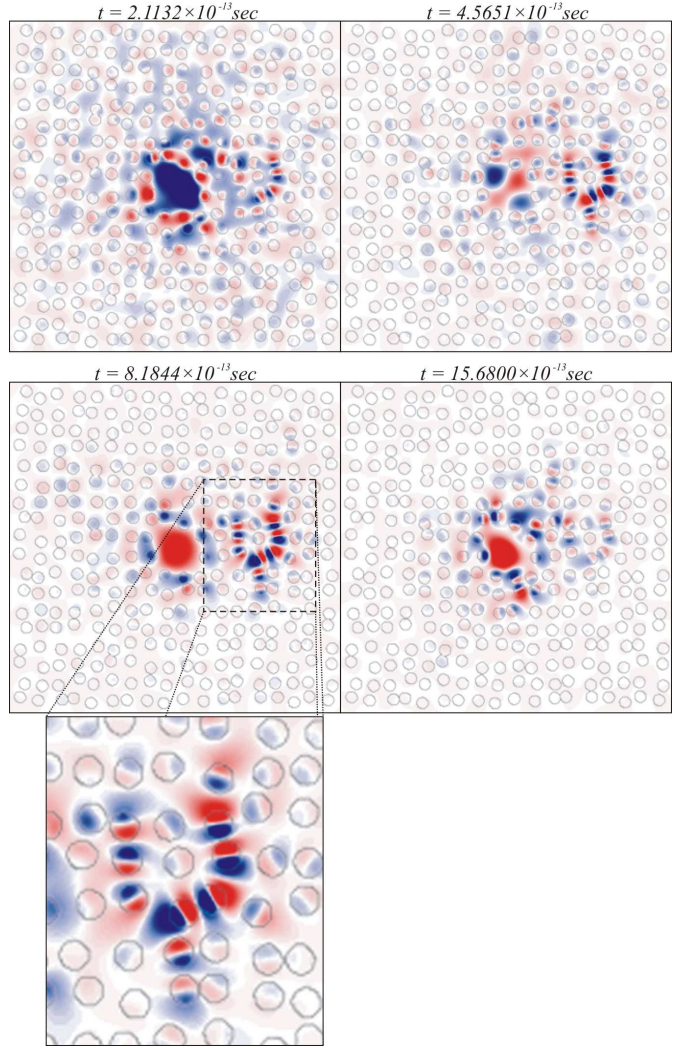}
    \caption{Snapshots of the field in a sample with $\alpha = 0.5$ at different times (blue for positive and red for negative values). Semi-localized mode magnified below. For real-time evolution see \cite{anim}}
    \label{fig:fig5}
\end{figure}
As can be seen from Fig.~\ref{fig:fig5}, increasing randomness raises the possibility of creation of semi-localized modes. Here, in addition to the central cavity localized mode, some other modes are confined and remain active inside the PC (see the {\textsf{\scalebox{1}[.6]{\footnotesize I}\hspace{-0.93ex}\raisebox{0.75ex}{\scalebox{1}[.65]{\footnotesize U}}\hspace{0.5ex}}}\normalfont shape pattern pulsating on the right side of the center in Fig.~\ref{fig:fig5}). This effect is due to the repeated internal reflections inside the randomly arranged cylinders which are somehow coupled or \emph{synchronized}. This synchronization occurs in a cluster of nano-particles whose (random) geometry allows for such coupling effect to take place. Since reflections are not total, these semi-localized modes weaken as time elapses. For a real-time dynamics showing the creation, evolution, and eventual dissipation of such localized modes see \cite{anim}.
Note that increasing randomness allows for more suitable situations to be created for semi-localized modes, and increasing the number of these possible modes broadens the output curve as seen in Fig.~\ref{fig:fig3} and also cause the maximum to occur later. Also, it is now easy to understand the lowering of the minima in Fig.~\ref{fig:fig4}: creation of semi-localized modes temporarily traps certain modes thus lowering the output energy in short times, releasing it later over a broad range of time.

\begin{figure}[htbp!]
    \includegraphics[width=\columnwidth]{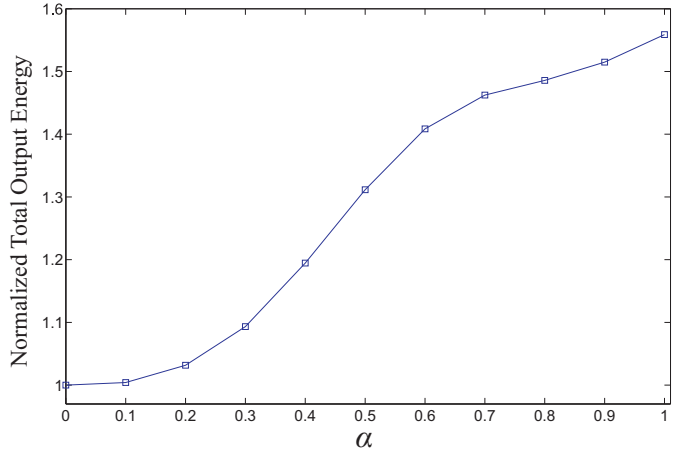}
    \caption{Total output energy from the crystal, normalized to the total output of ordered case, versus randomness factor.}
    \label{fig:fig6}
\end{figure}

\begin{figure}[htbp!]
    \includegraphics[width=\columnwidth]{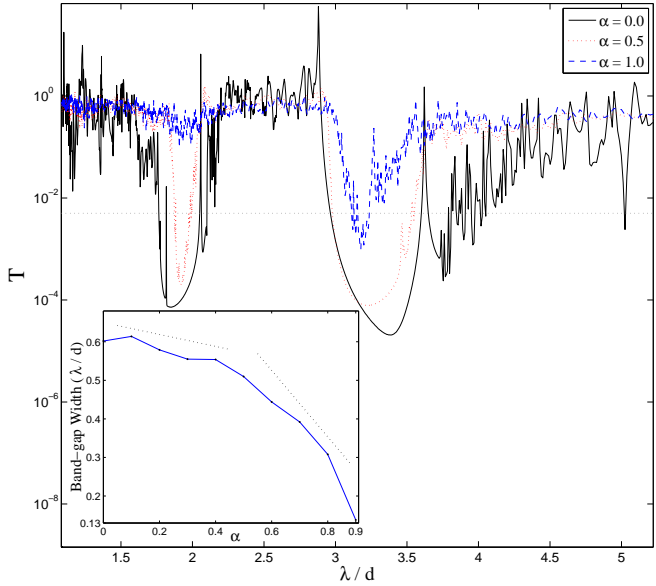}
    \caption{Transmission of PC for three different randomness factors as a function of normalized wavelength ($\lambda/d$). The inset shows the shrinking of the main band-gap as a function of randomness. Note the sudden change of behavior of $\alpha \approx 0.5$. Dotted lines are used to guide the eye.}
    \label{fig:fig7}
\end{figure}

How is the total energy output affected by increasing randomness? The answer is shown in Fig.~\ref{fig:fig6} where increasing randomness increases the total output energy. As pointed out above, this is due to the shrinking of the photonic band-gap and the resulting effect which allows for more modes to escape the PC. The shrinking of the photonic band-gap as a result of disorder can be best seen in Fig.~\ref{fig:fig7} where the transmission, $T=\l(\frac{E_{output}}{E_{input}}\r)^2$, has been plotted as a function of wavelength ($\lambda/d$) for three different randomness factors. We have actually plotted such figures for various values of $\alpha$ and note that the general structure of the primary (wider) band-gap remains intact up to $\alpha\approx0.5$ and steadily shrinks for larger values of $\alpha$. In the secondary (narrower) band-gap the observed effect of randomness is stronger, however the above-mentioned general scheme still holds.  The inset of Fig.~\ref{fig:fig7} shows in a more concrete way the relative robustness of the band-gap for small and intermediate $\alpha$, and its quick shrinking for larger $\alpha$.

\begin{figure}[htbp!]
    \includegraphics[width=\columnwidth]{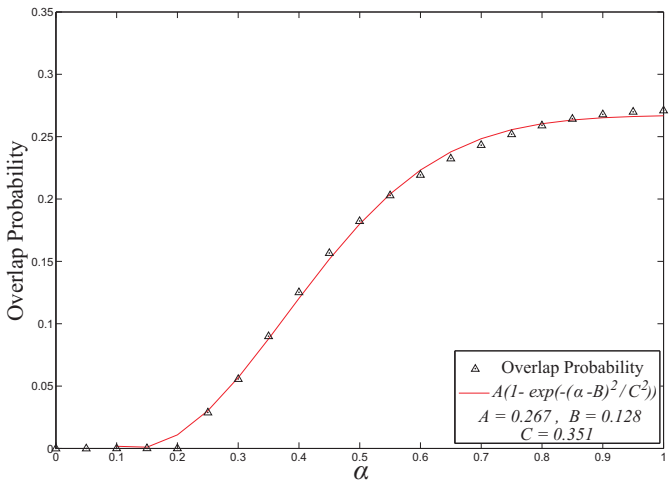}
    \caption{Overlap probability of a cylinder with its eight neighbors versus randomness factor. The solid line represents an exponential fit with the shown parameters in the inset.}
    \label{fig:fig8}
\end{figure}

Creation of semi-localized modes is a direct result of coupling of scatterers, so by calculating the probability of coupling of cylinders we can better answer the question of how randomness affects the propagation of electromagnetic waves. In our model, coupling happens when two cylinders are placed such that their center to center distance is less than or equal to the diameter ($2r$). We therefore calculate the probability of coupling of a cylinder with its eight nearest neighbor in a crystal for a given $\alpha$ and average over different realizations of the crystal (ensemble averaged). The result is plotted in Fig.~\ref{fig:fig8}.

The shape of this curve is very interesting and we believe underlies many of the observed behaviors in our study. Two simultaneous but distinct phenomena occur as randomness is increased~\cite{Sheng2006}: First is the shrinking and eventual destruction of the band-gap structure. Second is the creation of semi-localized modes. The correlation between the general shape of Fig.~\ref{fig:fig8} and many of our results (e.g. Figs.~\ref{fig:fig2},~\ref{fig:fig4},~\ref{fig:fig6},~\ref{fig:fig7}) is indicative of this general behavior. In Fig.~\ref{fig:fig8}, as $\alpha$ is increased from zero no significant change occurs, but a sharp (exponential) increase in the probability follows in the intermediate $\alpha$ range followed by a leveling off of the probability for larger randomness. This is precisely what is observed in Fig.~\ref{fig:fig2} and Fig.~\ref{fig:fig7} (due to destruction of band-gap structure), in Fig.~\ref{fig:fig4} (due to creation of semi-localized modes) and Fig.~\ref{fig:fig6} (due to both). Finally, we note that an inverse (adjusted) Gaussian fits well to the obtained data in Fig.~\ref{fig:fig8}, as indicated. The Gaussian shape of this behavior perhaps underlies the observed Gaussian behavior in Fig.~\ref{fig:fig2}.

\section{Conclusion}\label{sec4}
In this paper we have studied the transmissional properties of random two dimensional photonic crystals using FDTD method, in two experimentally relevant scenarios. First we study the effect of randomness in a $90\degr$ bent wave-guide and secondly we study transmissions along the edges of a centrally pulsed photonic crystal. By plotting various energy (and field) diagrams as functions of time and randomness, we have identified various interesting effects in the output of such samples. We study these properties in a full range of randomness and identify two important  effects: (i) the creation of semi-localized (or Anderson localized) modes  and, (ii) the shrinking of photonic band-gap. Semi-localized modes have been shown to arise as a result of synchronization of internal reflection modes in a random cluster of nano-particles whose close vicinity provides the required coupling. Furthermore, by correlating these two effects to overlap probability of nano-particles, we have identified the main cause of such effects. In this view the sudden rise in overlap probability around $\alpha \approx 0.5$ which underlies a sudden change in the band-gap structure is seen to explain many of our results. We suggest that experiments along our study be done in order to provide experimental verification of our results.

Finally, it has been shown that there is a correspondence in the active and passive modes in random media~\cite{PhysRevLett.87.183903,PhysRevA.69.031803}. Therefore, a possible extension of our present study is to introduce gain in our PC. Here, one would expect that the semi-localized modes studied here would also play a crucial role there. This might have important implications for the study of random lasers.


\end{document}